\documentclass[journal]{IEEEtran} 
\usepackage[utf8]{inputenc}
\usepackage{amsmath,graphicx,color,cite,amssymb,ulem,array}
\usepackage[dvipsnames]{xcolor}
\usepackage{soul}
\usepackage{authblk}

\title{Automated Screening for Distress: A Perspective for the Future}
\author[1]{Rajib	Rana}
\author[1]{Siddique Latif}
\author[1]{Raj Gururajan}
\author[1]{Anthony Gray}
\author[1]{Geraldine Mackenzie}
\author[3]{Gerald Humphris}
\author[1,4,5]{Jeff	Dunn}

\affil[1]{University of Southern Queensland, Australia}
\affil[3]{University of	St Andrews,	United Kingdom}
\affil[4]{Griffith University, Australia}
\affil[5]{University of Technology Sydney, Australia}

\begin{document}

\maketitle

\begin{abstract}
Distress is a complex condition which affects a significant percentage of cancer patients and may lead to depression, anxiety, sadness, suicide and other forms of psychological morbidity.  Compelling evidence supports screening for distress as a means of facilitating early intervention and subsequent improvements in psychological well-being and overall quality of life.  Nevertheless, despite the existence of evidence based and easily administered screening tools, for example the Distress Thermometer, routine screening for distress is yet to achieve widespread implementation.  Efforts are intensifying to utilise innovative, cost effective methods now available through emerging technologies in the informatics and computational arenas.
\end{abstract}
\section{Introduction}
Distress is described as emotional suffering with high global prevalence, which can result in disabling conditions and impairment for patients. 
%to live normal daily life. 
% It is a major health problem,  both in developing and developed nations, accounting for one-third of disability adjusted life years (DALYs) \cite{dachew2015prevalence}. 
It is highly prevalent in cancer patients affecting  25 to 60\% of patients~\cite{carlson2012screening} and can cause severe harm to this cohort with diminished Quality of Life (QoL) is one of the key adverse effects of distress~\cite{carlson2012screening}. Other implications include shortened survival time~\cite{brown2003psychological} and negative outcomes  for physical health through impaired immune functioning~\cite{levy1985prognostic}. Distress may also negatively influence treatment adherence in cancer patients and non-adherence substantially increases healthcare costs through increased likelihood of recurrence and other disease complications~\cite{demark2000current}. Overall, distress is estimated to increase the cost of cancer care by as much as 20\%\cite{chiles1999impact,otto2018assessing}.

Early assessment and screening for distress will enable timely management of distress leading to (1) improved adherence to treatment, (2) more effective communication between patient and clinician, (3) fewer visits to the hospital, and (4) early intervention for and prevention of severe anxiety or depression~\cite{national2003distress}. This is why International organisations have endorsed the quality care standard of whole-patient care that is achieved through routine comprehensive distress screening~\cite{lazenby2014international}, however, routine screening for distress has yet to be widely adopted~\cite{holland2011ipos}. Reasons for limited system wide update of screening for distress are reported to include  time constraints, limitations in the skills of health care providers, cost, and attitudes of health care providers to the use of standardised tools~\cite{carlson2012screening,chiles1999impact,lazenby2014international,holland2011ipos}. 

% In addition, most of these tools are primarily focused on post distress conditions such as depression, anxiety, Post-traumatic stress disorder (PTSD), and suicidal behaviour identification.  This trend needs to be re-imagined because if distress can be detected in the earlier stage, the consequences (depression, anxiety, suicidal ideation) might not happen in the first place.  

Automating  screening for distress with the assistance of emerging technologies may serve to alleviate many of the challenges associated with its routine applications. However, there are also a number of challenges need to be addressed to develop a robust automated distress detection system. In this paper, we aim to provide critiques on distress detection research in terms of assessment tools, available datasets and existing methods for automatic distress screening. A number of studies \cite{pampouchidou2017automatic,cummins2015review,morales2017cross} have reviewed the research on automatic depression detection but none of them has highlighted the research gap for distress detection. This study attempts to highlight the core differences of distress from other emotions and mental disorders, discuss existing methodologies for inferring distress and provide future directions for developing an automated distress system.
%by reviewing different studies on distress. 

% Almost every individual in their life have to face stressful event that exceed their abilities to cope with them. These stressful events can lead to long lasting distress whose detection is very important provide needed interventions.  Earlier distress detection systems can help people to avoid future consequences by detecting distress at an early stage. It is likely that when and if earlier distress detection systems are deployed, people will be able to get timely responses on their distress related illness. For timely distress detection, utilisation of technology based automatic solutions are ideal choice. Such automatic systems provide numerous advantages including large-scale and remote
% assessment, real-time identification of distress, continuous monitoring, complement traditional assessment methods, and also help professionals to manage patients more efficiently. 

% Technological advancement has made researcher to study mental health illnesses using automatic systems. Most of these attempts are primarily focused on post distress conditions such as depression, anxiety, Post-traumatic stress disorder (PTSD), and suicidal behaviour identification.  This trend need to be re-think because if distress can be detection in earlier stage then its consequences (depression, anxiety, suicidal ideation) can be tackled more effectively.  I

\section{What is Distress}
 Distress is considered as a continuum of psychological symptoms with varying severity \cite{carlson2003cancer}. In the forensic sciences the term ``distress'' is specifically used for the affective states that arise in violent situations. The National Comprehensive Cancer Network provides the widely accepted definition of distress\cite{national2003distress}: \textit{Distress is a multifactorial unpleasant emotional experience of a psychological (cognitive, behavioural, emotional), social, and/or spiritual nature. Distress encompass a range of common feelings of vulnerability, sadness, and fears that can cause depression, anxiety, panic, social isolation, and existential and spiritual crisis.} Distress may not always be caused by some unexpected external events, but can also be caused by internal states such as feelings, thoughts, and habitual behaviours \cite{linehan2018cognitive,leyro2010distress}. It is an uncomfortable feeling and can impact individuals' capacity of working, social life, bodily part and the mind. 
% People of all ages, gender, and creed having any economic status are vulnerable to distress.  
It is a subjective experience, and different people manifest it differently with varied range of symptoms. However, the most common symptoms \cite{Symptoms} of distress are:  sleep disturbances, memory problems, anger management issues, obsessive thoughts, fatigue, sadness, weight gain, hallucinations, delusions, etc.

\subsection{How Distress is different from Emotions}
Emotion is an essential component of human life and plays important role for their survival \cite{darwin1998expression}. As a human being, we feel a whole range of emotions that may be comfortable or uncomfortable \cite{ekman1992argument,ekman1971universals}. Emotional discomfort is a universal human experience. In fact, negative emotions including sadness, anger and fear are important and useful in various situations. For instance, fear is helpful when there is real threat to our safety (for e.g., gun pointed at us or wild ferocious animal in the vicinity) and helps humans to effectively withstand such threatening situations. Similarly, sadness inadvertently helps in spotlighting the things that we care about in our life and it is important reinforce that negative emotions are not necessarily distress \cite{emotion}, for example: disgust \cite{davey2011disgust}.

In our daily life, emotions are transient \cite{rana2016context} and they fluctuate like waves as they plateau,  subside and eventually pass. In other words, emotions are transient, continually  moving and changing. In contrast, distress is a prevailing situation that, if not addressed, escalates  until emotional combust~\cite{emotion}. Emotions such as fear or anger are aroused to prevent, solve, cope with, or get away from specific situation. Distress is different, it can be felt strongly, it compromises a person’s ability to cope and if left untreated may escalate to more serious conditions~\cite{distressli}. 
% In distress people often look for someone or call out for help, even scream.    

\subsection{How Distress is different from Stress}

Often stress and distress are used interchangeably, which blurs and confuses the distinctions between these concepts. It is however important to distinguish these two terms.  Stress is an important element of life, as it has both positive and negative effect. As pointed by Spielberger \cite{spielberger1987stress}, \textit{``Stress is an integral part of the natural fabric of life, and coping with stress is an everyday requirement for normal human growth and development''}. The body uses behavioural or physiological mechanisms to counter the perturbation caused by stress and come back to normality. People usually adapt stress but when this adaptive process is compromised stress  may develop into distress. Stress may present as either chronic and acute \cite{carstens2000recognizing} and any transition of stress to distress depends on various factors including duration, intensity, and controllability. 

\subsection{How Distress is different from Depression}
Other dimensions of psychopathology such as depression and anxiety are also closely related to distress. In particular, most assessment tools and the consequent treatment of distress is based on the depression symptoms \cite{fisher2007clinical}. Patients with depression need to meet at least five of the DSM-5 (Diagnostic and Statistical Manual of Mental Disorder, Fifth Edition) criterion for major depressive disorder nearly every day during a two-week period. However, distress has different symptoms such as poor self-management, feeling angry and scared, and feeling of unsupported by family and friends \cite{polonsky2005assessing}, which are not included in DSM-5. This suggests the need to formulate an alternative screening for distressed people, who are not clinically depressed. 

\begin{table*}[!ht]
\centering
\caption{Distress Scales}
\begin{tabular}{|m{3cm}|m{2.8cm}|m{2.5cm}|m{1.9cm}|}
\hline
\textbf{Scale}
&\textbf{Symptoms}
&\textbf{Number of Items}
&\textbf{Required Time (minutes)}
\\ \hline
\begin{tabular}[c]{@{}l@{}}BSI \cite{derogatis1983brief}\end{tabular}
&\begin{tabular}[c]{@{}l@{}}Psychological distress\end{tabular}
&\begin{tabular}[c]{@{}l@{}}18/53\end{tabular}
&\begin{tabular}[c]{@{}l@{}}3–7\end{tabular}
\\\hline

\begin{tabular}[c]{@{}l@{}}GHQ-12 \cite{goldberg1997validity}\end{tabular}
&\begin{tabular}[c]{@{}l@{}}Psychological disorders\end{tabular}
&\begin{tabular}[c]{@{}l@{}}12\end{tabular}
&\begin{tabular}[c]{@{}l@{}}3–7\end{tabular}
\\\hline

\begin{tabular}[c]{@{}l@{}}FACT-G \cite{webster2003f}\end{tabular}
&\begin{tabular}[c]{@{}l@{}}Quality of life\end{tabular}
&\begin{tabular}[c]{@{}l@{}}21\end{tabular}
&\begin{tabular}[c]{@{}l@{}}$<15$\end{tabular}
\\\hline

\begin{tabular}[c]{@{}l@{}}Distress Thermometer \cite{national2016nccn}\end{tabular}
&\begin{tabular}[c]{@{}l@{}}Distress\end{tabular}
&\begin{tabular}[c]{@{}l@{}}37\end{tabular}
&\begin{tabular}[c]{@{}l@{}}---\end{tabular}
\\\hline

\begin{tabular}[c]{@{}l@{}}SCL-90 \cite{derogatis1979symptom}\end{tabular}
&\begin{tabular}[c]{@{}l@{}}Psychiatric disorders\end{tabular}
&\begin{tabular}[c]{@{}l@{}}90\end{tabular}
&\begin{tabular}[c]{@{}l@{}}$>$20\end{tabular}
\\\hline

\begin{tabular}[c]{@{}l@{}}BDI \cite{olaya2010comparison}\end{tabular}
&\begin{tabular}[c]{@{}l@{}}Depression\end{tabular}
&\begin{tabular}[c]{@{}l@{}}21\end{tabular}
&\begin{tabular}[c]{@{}l@{}}5–10\end{tabular}
\\\hline

\begin{tabular}[c]{@{}l@{}}HADS \cite{zigmond1983hospital}\end{tabular}
&\begin{tabular}[c]{@{}l@{}}Depression/Anxiety\end{tabular}
&\begin{tabular}[c]{@{}l@{}}14\end{tabular}
&\begin{tabular}[c]{@{}l@{}}2-5\end{tabular}
\\\hline

% \begin{tabular}[c]{@{}l@{}}QIDS\end{tabular}
% &\begin{tabular}[c]{@{}l@{}}Depression\end{tabular}
% &\begin{tabular}[c]{@{}l@{}}16\end{tabular}
% &\begin{tabular}[c]{@{}l@{}}5–10\end{tabular}
% \\\hline

\begin{tabular}[c]{@{}l@{}}PHQ-9 \cite{williams2005performance}\end{tabular}
&\begin{tabular}[c]{@{}l@{}}Depression\end{tabular}
&\begin{tabular}[c]{@{}l@{}}9\end{tabular}
&\begin{tabular}[c]{@{}l@{}}$<$5\end{tabular}
\\\hline

% \begin{tabular}[c]{@{}l@{}}STAI \cite{speilberger1983state}\end{tabular}
% &\begin{tabular}[c]{@{}l@{}}Anxiety\end{tabular}
% &\begin{tabular}[c]{@{}l@{}}40\end{tabular}
% &\begin{tabular}[c]{@{}l@{}}5–10\end{tabular}
% \\\hline
\end{tabular}
\label{table:tools}
\end{table*}

 %General Health Questionnaire (GHQ-12)
%\subsection{Patient Health Questionnaire Eight-Item Depression Scale (PHQ-8)}

\section{Assessment Scales}

Distress remains undetected in most patients~\cite{holland2011ipos}, however, surprisingly, there are many scales available to gauge distress. 
% Due to the business in the oncological practice, clinicians are often reluctant to use the scales~\cite{mitchell2008acceptability}. 
In this section we present the most popular scales used to screen distress. We also present (see Table~\ref{table:tools}) the number of questions/items in each scale and time to conduct the screening to indicate the complexity of each scale. 
% There are a number of scales that have been built to gauge distress. 
% There are various tools that are used to scale distress. 
% The range of distress related markers is wide, however, some distress cues are found to be common among patients. 

The Disability Distress Assessment Tool (DisDAT) \cite{regnard2007understanding} was designed by a palliative care team to assess distress and is not a scoring tool, rather it documents a wide range of behaviours and signs related to distress. A distress scale based on ten symptoms was designed by Mccorkle et al. \cite{mccorkle1978development}. This scale was tested on 53 patients, where distress score was ranged from 10-41. The Distress Thermometer \cite{roth1998rapid}, is another scale which enables patients to rate their distress level on visual scale ranging from 0 (no distress) to 10 (extreme distress). The SCL-90 (Symptom Checklist-90) and BSI (Brief Symptom Inventory) have been widely used for screening of psychological distress in medical patients and demonstrated high levels of specificity and sensitivity \cite{zabora1990efficient,carlson2003cancer}. The 12-item General Health Questionnaire (GHQ-12) is designed to study of psychological disorders in general clinical setting and has been used in various studies \cite{kosidou2017trends,cuellar2014ghq}. A recently proposed scale for distress assessment is the K10 \cite{kessler2002short}. It is a a 10-item scale specifically designed to assess distress in population surveys. This scale evaluates the individuals on anxio-depressive symptoms over the last 30 days and provides a total score as an index of distress. The Functional Assessment of Chronic Illness Therapy (FACIT) Measurement System \cite{webster2003f} is used for the management of chronic illness using questionnaires related to health-related quality of life. Its generic version known as the Functional Assessment of Cancer Therapy-General (FACT-G) is compiled to use in four primary quality of life domains including physical well-being, social/family well-being, emotional well-being, and functional well-being. A six-item sub-scale of Somatic and Psychological Health Report (SPHERE-12) measures the aspects of distress and related conditions \cite{hickie2001development}. This scale is based on GHQ \cite{goldberg1997validity} and each item is scored on a three-point scale between 0 and 2, which gives a maximum score of 12.

%\subsection{Assessment tools for depression}

A number of scales for depression are also used to scale distress. Hospital Anxiety and Depression Scale (HADS) is a screening instrument that is used to assess anxiety and depression of physically ill patients \cite{zigmond1983hospital}. It includes 14 items for anxiety and depression with 4 alternative answers, which are used to measure total distress score. 
% The Hamilton Rating Scale for Depression (HAM-D) scale \cite{hamilton1960rating} has long been considered as the gold standard assessment tool for the diagnosis of depression. The HAM-D assessment provides a score on the level of depression based on different severity of symptoms including low mood, agitation, insomnia, anxiety and weight loss. It includes 21 questions with 3-5 possible responses and overall score is presented into five categories: normal (0–7), mild (8–13), moderate (14–18), severe (19–22) and very severe ($>$23). The reliability of both DSM and HAM-D, as diagnosis of Major Depressive Disorder (MDD), have been criticised \cite{bagby2004hamilton,chmielewski2015method}. Even the recent DSM-5 raises likelihood the false positive regarding diagnoses of depression by confusing it with sadness \cite{wakefield2015sadness}. Other diagnostic tools for depression rating includes: 16-item Quick Inventory of Depressive Symptomology (QIDS) \cite{rush200316},  10-item Montgomery–$\dot{A}$sberg Depression Rating Scale (MADRS) \cite{montgomery1979new}, and the 9-item Patient Health Questionnaire (PHQ-9) \cite{kroenke2001phq}. 
%These tools provide rating score depression and self-evaluated assessments offer great convenience compared to  clinician led assessments nut their reliability might be reduced due to patient reading ability and patient over-familiarity \cite{cusin2009rating}. 
% The diagnosis of depression can also be performed using self-report scales and inventories. Out of these These
Self-report scales including Beck’s Depression Inventory (BDI)~\cite{olaya2010comparison}, and Patient Health Questionnaire - Anxiety and Depression Scale (PHQ-ADS)~\cite{chilcot2018screening} have also been shown to have some relevance with distress for particular patient groups.

\section{Automatic Distress Assessment}
% Distress management involves the identification, monitoring, and prompt treatment of distress in all setting. It should be the integral part of medical care and patients should be screened at regular intervals, especially when disease status changes (i.e., remission, progression, recurrence, and clinical complications), for appropriate psychosocial assistance or any clinical care. 

Distress is highly prevalent in patients with chronic disease. Despite the fact that it can cause serious harm, clinicians are reluctant to use the existing distress screening for various reasons, most importantly for cost and time requirements \cite{mitchell2008acceptability}. Emerging information technologies are playing promising role to automate the screening of different health issues \cite{rana2016gait} and they also have great potential to be exploited for the automated screening of distress that may greatly alleviate these problems and facilitate widespread update. The potential benefit of automated screening for distress has encouraged research efforts and we discuss progress in this section.
Besides the application in health, automated distress detection has also been studied in two other areas and these are Aged Care and Forensics. In homes for elderly people distress calls arise if there is fall or a  fire or other such events \cite{aman2013speech}. In the forensic scenario, automated distress assists the Police prioritise the crime response based on the intensity of distress of the caller \cite{aihio2017improving}. Also, automated distress detection can  assist the forensic phoneticians by providing them an objective measure of distress of victims in recorded attacks. In this section, we  discuss the methodologies used in these three sectors.

\subsection{Health}

For automated distress detection in health, most of the studies focused on distress related conditions such as depression, anxiety, PTSD, and suicidal behaviour; very few studies \cite{pacula2014automatic,saleem2012automatic} have reported their results on distress detection. For instance,   an automated distress management system \cite{decker2016piloting} is piloted in outpatient medical oncology practice using tablet or computer for tailored psychosocial coping recommendations or referrals to individuals after immediate analysis. The authors used Distress Thermometer and problem list proposed by National Comprehensive Cancer Network as a screening tool. Their system matches patients identified concerns with the problem list and proposes evidence-based treatment suggestions and referrals. Verona coding definitions of emotional sequences (VR-CoDES) was developed for the detection and categorisation of patients' emotions and their corresponding healthcare physicians \cite{del2017verona}. Different studies have exploited VR-CoDES \cite{barracliffe2017can, del2011development,zhou2014applying}, however, the need for training of researchers on its usage and skilled labour necessary for labelling consultation recording are its major practical limitations. In this regard, Birkett et al. \cite{birkett2017towards} developed computer-based tools to assist VR-CoDES in the labelling of patients-physicians' recordings. The authors tried different representations of patients' utterances and evaluated well-known classifiers including  na\"ive  Bayes, logistic regressions, support vector machines, and boosted ensemble decision trees for the labelling of recordings as an  explicit  concern, an  emotional  cue, or neither.  %\RR{what is recordings in three expressions?}.

Researchers are predominantly attempting to infer distress based on the after effect such as depression, anxiety, PTSD, and suicidality, have developed various techniques. In \cite{girard2014nonverbal}, authors analysed 33 individuals from a clinical trial of depression \cite{cohn2009detecting} and investigated the relationship between nonverbal behaviour and severity of depression using video recording over the course of treatment. Scherer at al. \cite{scherer2013automatic} evaluated different visual features for psychological disorder analysis. They found that depressed individuals tend to gaze downwards more, give less intense and shorter duration of smile, and show longer self-touches and fidgeting. The inclusion of gender information with the visual is found to be helpful in detecting of distress related situations \cite{stratou2015automatic}. In addition to the visual indicators, Space-Time Interest Points (STIP) features are also exploited to detect depression with significantly improved results \cite{cummins2013diagnosis,joshi2013can}. These features include gestures related to head, face, shoulder, hands movements. 

Recent studies have shown the promise of using speech as an effective marker for diagnosis and monitoring of depression. Speech can provide a wide range of prosodic and spectral features that can be effectively being used for human emotion \cite{latif2017variational,latif2018transfer} and depression detection.  Many researchers have used speech as an objective indicator for the detection of depression \cite{cummins2011investigation,cummins2014variability,scherer2013investigating}. An interactive voice response (IVR) system was used to collect speech samples for automated HAM-D measures of depression severity \cite{kobak1999computerized,mundt1998administration,moore2006examination,mundt2007voice}. Acoustic features such as spectral, prosodic, cepstral, glottal, and features obtained from Teager energy operators (TEO) were investigated for clinical depression detection in adolescents \cite{low2011detection}. TEO based features were produced more promising results compared to all other features and their combinations. Other studies \cite{alpert2001reflections,darby1977vocal,france2000acoustical,moore2008critical} also investigated different acoustic features and identified more relevant identifier for depression. Ozdas et al. \cite{ozdas2004investigation} studied excitation related speech parameters including glottal flow spectrum and vocal jitter for identification of major depressed, high-risk near-term suicidal, and non-suicidal patients. Vocal jitter was found a significant discriminator clue suicidal and non-depressed control, where glottal flow spectrum related parameters provided discrimination of all three groups with significantly improved results. Scherer et al. \cite{scherer2013investigating} used prosody and voice quality related speech parameters for identification of suicidal and non-suicidal adolescents. They found that suicidal adolescents tend to have more breathy voice qualities compared to non-suicidal. A comparative study performed in \cite{alghowinem2013comparative} using acoustic and prosodic features to detect depression in spontaneous speech. Authors found that voice features such as intensity, root mean square, and loudness performed best to detect depression in the dataset.  Other studies (for example \cite{kandsberger2016using,salekin2018weakly,lopez2017depression,dham2017depression,scherer2015reduced}) also exploited different machine learning techniques and suggested that the speech can be effectively utilised to detect distress and related conditions.

%\RR{ and what did they find? if you do not add this there is no point of add these here. }. 

\subsection{Aged Care}
Life expectancy is increasing globally, leading us to a higher number of older people in our society \cite{latif2018mobile}. This increasing share of the elderly population is in part responsible for a shift in the cause of death from infectious and parasitic illnesses to chronic non-communicable diseases \cite{latif20175g,latif2017mobile}. Ageing can lead to physical limitations that need to be compensated by technical assistance and the help of aged care services. In aged care residential communities, feeling of isolation, fear, and a sense of helplessness, such as an inability to perform routine tasks, may lead to distress \cite{berardo1970survivorship,iliffe2007health}.

Distress in elderly people often goes unrecognised for a range of reasons including confusing or unknown symptoms of distress~\cite{shivakumar2015identifying}, avoidance from checkups \cite{davison2008improving}, and lack of systematic method or tool for distress detection \cite{clarke2002questionnaire}. The early detection and treatment of distress among elderly people is  important because it can enhance recovery from illness and improve overall quality of life~\cite{shivakumar2015identifying}. There exist different innovative products and solutions which promote independence and better quality of life among seniors with physical or cognitive diseases, for instance, the CIRDO project~\cite{bouakaz2014cirdo} aims to automatically detect the situations of falls and distress in residential care to promote autonomy for elderly people. This system involves video and audio analysis to detect the risky situation and make necessary emergency call using e-lio system\footnote{ http://www.technosens.fr/.}  For distress detection, CIRDO evaluated the proposed system using Automatic Speech Recognition (ASR) to detect distress sentences in AS80 \cite{aman2013speech} corpus and achieved promising results. The SweetHome project \cite{vacher2012recognition} used home equipped noise robust multisource automatic speech recognition (ASR) to detect vocal command or distress sentences in the realistic noisy environment of a smart home. Twenty three subjects or “speakers”, participated in this experiment where the closest distance between speakers and microphone was two meters. The authors performed voice order recognition of speech command belonging to three classes: distress calls, neutral sentences, and home automation orders. Alternatively, a sound based surveillance system~\cite{istrate2006information} to detect alarming sounds in home situation has been described. This system performed real-time audio analysis for the detection of distress situation without compromising patients' privacy.

Distress detection in elders using ASR system is a very challenging task due age-related degeneration of vocal cords, problems of laryngeal cartilages, and changes in larynx muscles \cite{takeda2000aging,mueller1985acoustic}. Some studies have empirically shown that ASR models performed poorly on elderly voice when they are trained on young or middle‐aged adult speech \cite{vipperla2008longitudinal,vigouroux2004etude}. For such situations, speaker adaptation techniques or training ASR model on elderly voice can help improvement in recognition rate \cite{baba2004acoustic}. To explore the performance of ASR in distress situation, Aman et al. \cite{aman2013speech} presented word error rate in aged voice compared to non-aged speech. They showed that ASR system gives higher word error rate equal to equal to 43.5\% for the aged group and 9\% on young speakers.  
% There is an imperative need of such systems that can improve distress detection for elderly people in aged care.  

\subsection{Forensic}

Distress detection has an increasing presence in forensics, particularly in informing opinions about the authenticity of distress in criminal investigations. In forensic investigation, the lie can occur from distortion, denial, evasion, concealment, and outright fabrication by people to appear non-accountable for their exertions \cite{frank2004nonverbal} and distress surveillance  systems are used to identify the presence of reliable emotional clues to detect malingering or deception. Forensic examinations are performed by psychologists using different techniques including interviews, observations, home or institutional visits, psychological tests and instruments, as well as other methods recognised by the Forensic Council \cite{gava2013techniques}. Automatic distress detection can play a crucial role in the assisting the forensic examination practice with an objective measure that can assist the judicial authorities.

A comprehensive study was performed by Lisa \cite{roberts2012forensic} to investigate distress in speech using acoustic and perceptual cues and empirically compare the results for real-life victims and actors in life-threatening situations. Based on the results of the acoustic analysis, it is concluded that acoustic parameters can be utilised to detect distress situations for actors and victims. In another study, Lisa \cite{roberts2011acoustic} reported that two acoustic parameters intensity and formant bandwidth are helpful in differentiating between acted and genuine victims' speech. Similarly,  Fundamental Frequency (F0) mean, range and vowel formant can be used to distinguish between baseline and distress conditions for both victims and actors.

\begin{table*}[!ht]
\centering
\caption{Studies on automated distress detection in different settings. %\RR{Can you find out 1. how many study subjects were used, 2. some description of the subjects,3. What accuracy or other measure of success etc. \\Give a bit more detail in the ``Method'', such as using ASR etc. Repeat this for each method. This should be a self-contained table. People should be able to get just enough idea about the landscape by reading this table.} 
}
\scriptsize
\begin{tabular}{ |m{1.1cm} | m{2.8cm} | m{2.5cm} | m{1.3cm} | m{2.5cm} |m{2.5cm} |m{2.2cm} |}
\hline
\textbf{Scope}
& \textbf{Corpus}
& \textbf{Scale}
& \textbf{Modality} 
& \textbf{Method} 
& \textbf{Focus} 
& \textbf{References}\\ \hline
%&\begin{tabular}[c]{@{}l@{}}Distess Sentences\end{tabular}  
&\begin{tabular}[c]{@{}l@{}}AD80\end{tabular}
%&\begin{tabular}[c]{@{}l@{}}Speech\end{tabular} 
%&\begin{tabular}[c]{@{}l@{}}Speech to text\end{tabular}
&&&&&\begin{tabular}[c]{@{}l@{}} \cite{aman2013speech,aman2013home,bouakaz2014cirdo,vacher2015recognition,aman2016influence,vacher2015development,aman2013analysing} \end{tabular} \\  \cline{7-7}\cline{2-2}
\begin{tabular}[c]{@{}l@{}}\textbf{Aged Care}\end{tabular}
&\begin{tabular}[c]{@{}l@{}}French Adapted Speech\\
Corpus\end{tabular}
&\begin{tabular}[c]{@{}l@{}}Distressed Sentences\end{tabular}                 
&\begin{tabular}[c]{@{}l@{}}Speech\end{tabular} 
&\begin{tabular}[c]{@{}l@{}}Speech to Text\\ using ASR\end{tabular}
&\begin{tabular}[c]{@{}l@{}}Distress Situation \\Detection\end{tabular}
&\begin{tabular}[c]{@{}l@{}} \cite{istrate2008embedded,vacher2006speech,fleury2008sound,vacher2009speech,vacher2016cirdo,vacher2010complete,vacher2010challenges} \end{tabular} \\  \cline{7-7}\cline{2-2}

%&\begin{tabular}[c]{@{}l@{}}Distress Sentences\end{tabular}   
&\begin{tabular}[c]{@{}l@{}}CIRDO Corpus\end{tabular}
%&\begin{tabular}[c]{@{}l@{}}Speech\end{tabular} 
%&\begin{tabular}[c]{@{}l@{}}Speech to text\end{tabular}
&&&&&\begin{tabular}[c]{@{}l@{}} \cite{vacher2016cirdo,vacher2015recognition,lecouteux2018distant} \end{tabular} \\  \cline{1-7}

\begin{tabular}[c]{@{}l@{}}\textbf{Forensic}\end{tabular}  
& \begin{tabular}[c]{@{}l@{}}Self Generated by Author\end{tabular}
&\begin{tabular}[c]{@{}l@{}}Distressed/non-distressed\\ speech\end{tabular}
& \begin{tabular}[c]{@{}l@{}}Speech\end{tabular} 
& \begin{tabular}[c]{@{}l@{}}Acoustic Analysis\end{tabular} 
& \begin{tabular}[c]{@{}l@{}}Distressed/non-distressed\\ speech detection\end{tabular} 
& \begin{tabular}[c]{@{}l@{}}\cite{roberts2012forensic,roberts2011acoustic}\end{tabular} 
\\ \cline{1-7}

&&\begin{tabular}[c]{@{}l@{}}PCL-C, PHQ-9\end{tabular}
%&\begin{tabular}[c]{@{}l@{}}Audio, Video\end{tabular}
%&&\begin{tabular}[c]{@{}l@{}}Verbal or Non-verbal\\ Analysis\end{tabular}
&&&\begin{tabular}[c]{@{}l@{}}Depression and PTSD\end{tabular}
&\begin{tabular}[c]{@{}l@{}} \cite{lucas2015towards,scherer2016self,scherer2013investigating} \end{tabular} \\ \cline{3-3}\cline{6-7}

&\begin{tabular}[c]{@{}l@{}}Distress Analysis Interview\\ Corpus (DAIC)\end{tabular}
&\begin{tabular}[c]{@{}l@{}}PHQ-9\end{tabular}
&\begin{tabular}[c]{@{}l@{}}Audio, Video\end{tabular}
&\begin{tabular}[c]{@{}l@{}}Verbal/Non-verbal/Fusion\\ Analysis\end{tabular}
&\begin{tabular}[c]{@{}l@{}}Depression\end{tabular}
&\begin{tabular}[c]{@{}l@{}} \cite{scherer2013audiovisual,yu2013multimodal,stasak2016depression} \end{tabular} \\ \cline{3-3}\cline{6-7}

&&\begin{tabular}[c]{@{}l@{}} STAI, PHQ-9, PCL-C \end{tabular}
%&\begin{tabular}[c]{@{}l@{}}Audio, Video\end{tabular}
%&\begin{tabular}[c]{@{}l@{}}Verbal or Non-verbal\\ Analysis\end{tabular}
&&&\begin{tabular}[c]{@{}l@{}}\textbf{Distress}\end{tabular}
&\begin{tabular}[c]{@{}l@{}} \cite{scherer2014automatic,scherer2013automatic} \end{tabular} \\ \cline{2-7}%\cline{3-7}

&\begin{tabular}[c]{@{}l@{}}Virtual Human Distress \\Assessment Interview \\Corpus (VH DAIC)\end{tabular}
&\begin{tabular}[c]{@{}l@{}}PHQ-9, PCL-C\end{tabular}
&\begin{tabular}[c]{@{}l@{}}Audio, Video\end{tabular}
&\begin{tabular}[c]{@{}l@{}}Verbal/Non-verbal/Fusion\\ Analysis\end{tabular}
&\begin{tabular}[c]{@{}l@{}}Depression, and PTSD\end{tabular}
&\begin{tabular}[c]{@{}l@{}} \cite{stratou2013automatic,chatterjee2014context,stratou2015automatic} \end{tabular} \\ \cline{2-7}

\begin{tabular}[c]{@{}l@{}}\textbf{Healthcare}\end{tabular}
&\begin{tabular}[c]{@{}l@{}}Distress Analysis Interview\\ Corpus-Wizard of Oz\\ (DAIC-WOZ)\end{tabular}
&\begin{tabular}[c]{@{}l@{}}PHQ-8\end{tabular}
&\begin{tabular}[c]{@{}l@{}}Audio, Video,\\ Text\end{tabular}
&\begin{tabular}[c]{@{}l@{}}Verbal/Non-verbal/Text/\\Fusion Analysis\end{tabular}
&\begin{tabular}[c]{@{}l@{}}Depression\end{tabular}
&\begin{tabular}[c]{@{}l@{}} \cite{yang2016decision,williamson2016detecting,nasir2016multimodal,pampouchidou2016depression,vlasenko2017implementing,cummins2017enhancing,yang2017dcnn,yang2017hybrid,yang2017multimodal,morales2018linguistically,song2018human,stasak2019investigation} \end{tabular} \\ \cline{2-7}%\cline{3-3}\cline{6-7}

&\begin{tabular}[c]{@{}l@{}}Audio-Visual Depressive \\Language (AViD) Corpus,\\AVEC 2014 Audio-Visual \\Depression Corpus (AVEC)\end{tabular}
&\begin{tabular}[c]{@{}l@{}}BDI-II\end{tabular}
&\begin{tabular}[c]{@{}l@{}}Audio, Video\end{tabular}
&\begin{tabular}[c]{@{}l@{}}Verbal/Non-verbal/\\Fusion Analysis\end{tabular}
&\begin{tabular}[c]{@{}l@{}}Depression\end{tabular}
&\begin{tabular}[c]{@{}l@{}} \cite{kachele2014fusion,meng2013depression,jan2014automatic,zhu2017automated,gupta2016predicting,ma2016cost,williamson2013vocal,meng2014automatic,al2018video,zhou2018visually,jan2017artificial} \cite{valstar2013avec} \cite{scherer2015reduced} \\\cite{cummins2013diagnosis,kaya2014cca,mitra2014sri}\\\cite{stasak2019investigation}\end{tabular} \\ \cline{2-7}%\cline{3-3}\cline{6-7}

&\begin{tabular}[c]{@{}l@{}}Black Dog\end{tabular}
&\begin{tabular}[c]{@{}l@{}}DSM-IV, HAM-D\end{tabular}
&\begin{tabular}[c]{@{}l@{}}Audio, Video\end{tabular}
&\begin{tabular}[c]{@{}l@{}}Verbal/Non-verbal/Fusion\\ Analysis\end{tabular}
&\begin{tabular}[c]{@{}l@{}}Depression\end{tabular}
&\begin{tabular}[c]{@{}l@{}} \cite{joshi2013can,joshi2012neural,joshi2013multimodal,alghowinem2013eye,alghowinem2013head,alghowinem2016multimodal,mcintyre2009approach,mcintyre2011facial} \end{tabular} \\ \cline{2-7}

&\begin{tabular}[c]{@{}l@{}}Pittsburgh\end{tabular}
&\begin{tabular}[c]{@{}l@{}}DSM-IV, HAM-D\end{tabular}
&\begin{tabular}[c]{@{}l@{}}Audio, Video\end{tabular}
&\begin{tabular}[c]{@{}l@{}}Verbal/Non-verbal/Fusion\\ Analysis\end{tabular}
&\begin{tabular}[c]{@{}l@{}}Depression\end{tabular}
&\begin{tabular}[c]{@{}l@{}}\cite{dibekliouglu2018dynamic,joshi2013automated,joshi2013relative,cohn2010social,girard2014nonverbal,girard2013social,cohn2013beyond,alghowinem2015cross,dibekliouglu2015multimodal} \\ \cite{cohn2009detecting}\end{tabular} \\ \cline{2-7}

&\begin{tabular}[c]{@{}l@{}}ORYGEN\end{tabular}
&\begin{tabular}[c]{@{}l@{}}Conventional diagnostic \\tests\end{tabular}
&\begin{tabular}[c]{@{}l@{}}Audio, Video\end{tabular}
&\begin{tabular}[c]{@{}l@{}}Verbal/Non-verbal/Fusion\\ Analysis\end{tabular}
&\begin{tabular}[c]{@{}l@{}}Depression\end{tabular}
&\begin{tabular}[c]{@{}l@{}} \cite{ooi2011prediction,ooi2014early} \end{tabular} \\ \cline{1-7}

\end{tabular}

\label{Sum}
\end{table*}

\section{Distress Datasets}
\label{data}
Development of automated systems require historical data that contains the correlation of physical properties such as speech, facial expressions with distress labels. In this section we discuss various datasets that have been used previously for the purpose of distress identification are identified and discussed. As depression is a possible after effect of distress most of these datasets are built to diagnose depression. This section concludes with a summary of the sector wise studies (health, age care, and forensic) with the datasets used within, in Table~\ref{Sum}.
% 

% can this also provide them the opportunity to decide on the   to Distress can cause enormous social and economic disadvantages to population and above 50\% of patients with distress related disorders are remain untreated by their primary care physicians \cite{higgins1994review}. For designing automated distress detection systems, data collection is a first and crucial step. There are few distress dataset that are discussed below. 

\subsection{Distress Assessment Interview Corpus (DAIC)}

The Distress Analysis Interview Corpus (DAIC) \cite{gratch2014distress} includes semi-structured clinical interviews of participants to enable the diagnosis of psychological distress conditions such as depression, anxiety, and post-traumatic stress disorder. The interviews of participants were conducted by humans, human controlled agents and autonomous agents. Overall data consists of audio, video, and questionnaire responses of participants and each interview is labelled with a depression score using PHQ-9. A portion of this dataset was released in Audio/Visual Emotion Recognition Depression Sub-challenge (AVEC) \cite{valstar2016avec} 2016, which also contains transcription of the interviews.

\subsection{Aged and Non-Aged Corpus (AS80)}

This corpus was recorded for adaptation of standard Automatic Speech Recognition (ASR) system to aged voice \cite{aman2013speech}. This corpus contains recording of from 95 speakers who were asked to read  distress and casual sentences. These sentences contain a list of home automation orders and of distress  calls that could be uttered by an elderly person in distress or fall situations. 

\subsection{AVEC Corpus 2013}
This dataset contains 340 video recordings of subjects performing a Human-Computer Interaction tasks \cite{valstar2013avec}. There were total 292 speakers and the length of each recorded video clip is between 20 to 50 minutes. The level of depression for recordings was labelled  using Beck Depression Inventory (BDI-II) \cite{beck1996beck}. The AVEC corpus 2014 \cite{valstar2014avec} is a portion of this dataset which contains 300 video with the duration from 6 seconds to 4 minutes.

\subsection{SDC (suicidal, depressed, and control subjects)}
This database is the collection of different dataset. Suicidal corpus was collected from the existing datasets \cite{silverman2006methods} that was recorded from phone conversations, treatment sessions, and suicide notes. Depression related samples were obtained from Vanderbilt II and depression dataset used by Hollon et al \cite{hollon1992cognitive}. DSM-IV and, ICD-9-CM (International Classification of Diseases, ninth edition, Clinical Modification) criteria were used for depressed patients. For the control group sample, Vanderbilt II dataset was used.

\subsection{Pitt Depression Dataset}

This is a clinically validated depression dataset collected during the treatment of depressed patients at University of Pittsburgh (Pitt) \cite{yang2013detecting}. All participants from a clinical trial were met with DSM-IV criteria for major depression. Total 57 patients were accessed using the HRSD clinical interview for depression severity. Interviews were recorded in audio-video format and depression was evaluated by the clinicians.

\subsection{Black Dog dataset}
This audio-visual dataset was recorded by the Black Dog Institute Australia \cite{alghowinem2012joyous}. Over 40 depressed individuals (both male and female) were interviewed and asked to read sentences. Audio-video recordings of subjects include self-directed speech, related facial expressions, and body language.

\subsection{Cincinnati Children’s Interview Corpus (CCIC)}
This dataset \cite{scherer2013investigating} includes the interview of 60 children patients (average age 15.47 years) at the Emergency Department of Cincinnati Children’s Hospital Medical Center. These children came to the hospital due to suicidal ideation, gestures, and attempts. Data was collected by a professional social worker. Due to lengthy interviews of suicidal and non-suicidal patients, only 60 seconds of speech for each participant is utilised for the analysis \cite{venek2014adolescent}.

\section{Discussions}

A search of the  literature for research focused on the application of automation in screening for distress found most of the papers  in the health area utilizing  post-distress conditions including depression, anxiety, and Post Traumatic Stress Disorder (PTSD) as a proxy to determine the presence of distress, rather than screen for distress itself. More broadly, it is evident from Table~\ref{Sum} where a summary of 75 studies relating to   automated approaches to screening for distress in the aged care, forensic and health care settings is presented, that the focus is mostly on depression and PTSD. These studies have statistically analysed depression, anxiety, and PTSD against distress and reported that distress is highly correlated with these measuring dimensions~\cite{bieling1998state,marshall2010all,arbisi2012predictive}. Only two studies were found, refer Table \ref{Sum},  \cite{scherer2013automatic,scherer2014automatic} targeted distress specifically using automated methods. However, even these two studies statistically correlated depression, anxiety, and PTSD to categorise (high, low, unclear) distress. In cancer care, this approach, where other conditions such as depression, are used as a proxy or signal for distress has limitations as as  patients with distress may not have depression when measured with the existing scales \cite{tate1993prevalence} and as such may not be detected. Future research needs to focus on automated approaches to screening for distress in cancer patients which are independent from other related conditions (such as anxiety and depression) and are designed specifically to identify symptoms associated with distress. Guidance can be found in promising work  from the  Forensic~\cite{roberts2011acoustic} and Health~\cite{del2017verona} areas where it has been shown that  speech independently carries latent properties for inferring distress. 

Existing tools to screen for distress vary in relation to complexity, have been criticised as being costly, in terms of time and resources, and have failed to attract widespread or routine implementation, refer Table~\ref{table:tools}. Evidence based automated approaches which efficiently and effectively screen for distress without adding to patient or staff burden may well be the future of screening for distress.  Such an approach would triage high distress individuals for the attention of professional staff for further assessment or referral, consistent with a tiered model of care approach~\cite{hutchison2006tiered}. 

Currently, a number of distress screening tools are available, but as we report in Table~\ref{table:tools}, most of these tools have many questions and require a considerable amount of time to complete. More importantly, it has been found that screening patients with multiple scales can appreciably improve the accuracy of results compared to single scale \cite{clover2009my}. However, such multi-scale approach will further increase the screening time. Due to busy practices, oncologists are already reluctant to use distress screening tools, so a further increase in screen time will not be welcome by the oncology practices. Moreover, for aged care, and forensic scenarios, real-time distress inference is sought, so the time taking screening techniques will not be very useful. {\bf An automated distress detection/screening is therefore inevitable.}

Datasets will play a vital role to develop an independent and automated distress detection system. From Section \ref{data}, we found that most of the available datasets are recorded for depression, anxiety, and assessment of suicidal behaviour. Very few datasets such as AD80 is designed for distress in the elderly population. However, AD80 only focuses on the detection of distressed sentences using Automated Speech Recognition that focus on words (``help'') spoken by individuals. In addition, each dataset has been recorded in different environment and validated using different scales, which is hard to use for developing the automated distress detection tool. {\bf Therefore, it is crucial to develop large-scale validated datasets for automated detection of distress.}

\end{document}